\documentclass{article}

\usepackage{arxiv}

\usepackage[utf8]{inputenc} 
\usepackage[T1]{fontenc}    
\usepackage{hyperref}       
\usepackage{url}            
\usepackage{booktabs}       
\usepackage{amsfonts}       
\usepackage{nicefrac}       
\usepackage{microtype}      
\usepackage{lipsum}
\usepackage{graphicx}
\graphicspath{ {./images/} }
\usepackage{subcaption}
\usepackage{amsmath} 

\title{Pulmonary Tuberculosis Edge Diagnosis System Based on MindSpore Framework: Low - cost and High - precision Implementation with Ascend 310 Chip }

\author{
 HaoYu Li \\
  Changchun University Of Finance And Economics\\
}

\begin{document}
\maketitle
\begin{abstract}
Pulmonary Tuberculosis (PTB) remains a major challenge for global health, especially in areas with poor medical resources, where access to specialized medical knowledge and diagnostic tools is limited. This paper presents an auxiliary diagnosis system for pulmonary tuberculosis based on Huawei MindSpore framework and Ascend310 edge computing chip. Using MobileNetV3 architecture and Softmax cross entropy loss function with momentum optimizer. The system operates with FP16 hybrid accuracy on the Orange pie AIPro (Atlas 200 DK) edge device and performs well. In the test set containing 4148 chest images, the model accuracy reached 99.1\% (AUC = 0.99), and the equipment cost was controlled within \$150, providing affordable AI-assisted diagnosis scheme for primary care. 
\end{abstract}


\section{Introduction}
PTB (Pulmonary Tuberculosis), a chronic infectious disease caused by Mycobacterium tuberculosis (M.tb), remains a serious public health issue that threatens human health globally, as highlighted in the World Health Organization’s (WHO) 2024 report, Global Tuberculosis Report 2024 . Regions with high incidence of tuberculosis often face economic underdevelopment and poor medical conditions, lacking both professional healthcare workers and advanced diagnostic equipment. This makes early diagnosis and treatment of tuberculosis particularly challenging. Traditional X-ray-based diagnosis of pulmonary tuberculosis relies heavily on doctors' expertise and visual observation, which is not only inefficient but also prone to subjective influences, leading to higher risks of misdiagnosis and missed diagnoses. In impoverished areas with scarce medical resources or underdeveloped healthcare systems, these limitations are even more pronounced. Therefore, developing a low-cost, high-accuracy auxiliary diagnostic solution for tuberculosis using X-rays should be an urgent priority.

In recent years, the application of artificial intelligence (AI) technology in the medical field has provided new possibilities for addressing these challenges \cite{sun2024explainableartificialintelligencemedical}. Particularly, the outstanding performance of deep learning algorithms in image recognition and classification tasks has made them ideal tools for assisting in medical imaging diagnostics. Research has shown that models based on convolutional neural networks (CNNs) can efficiently process complex medical imaging data \cite{Qayyum2017MedicalIA}\cite{Rahman2020ReliableTD}, and multiple studies have demonstrated that AI-assisted human diagnosis can improve diagnostic speed \cite{chen2025domainexpertsrelyai}. However, as model accuracy improves, their depth typically increases, resulting in more complex model architectures \cite{he2015deepresiduallearningimage}\cite{simonyan2015deepconvolutionalnetworkslargescale}. The main obstacle to applying these advanced AI technologies in resource-limited regions lies in the high computational costs and hardware requirements.

To address this issue, this study proposes a deployment solution for a low-cost, high-performance edge computing device based on the Ascend 310 chip. The solution utilizes a lightweight yet highly accurate convolutional neural network model, which is applied to assist in diagnosing tuberculosis from X-ray images in regions with underdeveloped healthcare infrastructure.

\section{Background and Methodology}
\label{sec:headings}
At present, most researchers are exploring how to apply artificial intelligence (AI) to our daily lives in order to solve practical problems. Particularly in the medical field, there is a strong focus on how AI can assist doctors in making rapid diagnoses, thereby improving diagnostic efficiency. Currently, edge computing devices are widely used to deploy AI systems at a low cost in grassroots healthcare environments. However, many of the commonly used edge computing devices, such as NVIDIA’s Jetson series \cite{basit2024medaideleveraginglargelanguage}\cite{mir2024democratizingmllmshealthcaretinyllavamed}\cite{s2024sepipelinevisiontransformer}, tend to be relatively expensive and do not offer competitive memory or computational power compared to other devices in the same price range.

In contrast, within the same price bracket, we have chosen the Orange Pi AIPro, which offers higher computational power and memory, allowing for more efficient and convenient deployment of AI solutions. The Orange Pi AIPro is equipped with Huawei's Ascend 310 AI chip, and the MindSpore deep learning framework provides better support for the Ascend series of chips. Theoretically, this setup allows for higher accuracy with lower energy consumption, making it a more effective solution for deploying AI in resource-constrained healthcare environments.

\subsection{Ascend 310 Chip}
The Ascend 310 chip is an AI computing chip released by Huawei in 2018. With a maximum power consumption of just 8W, it is highly suitable for edge computing scenarios that require high energy efficiency. In our hardware device, the Orange Pi AIPro, this chip supports half-precision (FP16) computing at 6 TFLOPS and integer precision (INT8) computing at 12 TOPS. Additionally, it is capable of supporting multi-channel full HD video decoding.

\subsection{MindSpore Deep Learning Framework}
MindSpore is an open-source AI framework introduced by Huawei, designed to support flexible deployment across cloud, edge, and device scenarios. In this research, MindSpore played a significant role in unleashing the computational power of the hardware. By using MindSpore, we were able to fully harness the computing capabilities of the Ascend 310 chip, significantly reducing memory usage while accelerating computation speed.

\subsection{MobileNet Model}
MobileNet is a series of highly efficient and lightweight convolutional neural network models proposed by Google, designed to provide efficient computational capabilities for mobile and embedded devices. The requirements of this study demand that the model be both lightweight and highly accurate. Ultimately, the latest MobileNetV3 model was chosen for this purpose. Compared to its predecessors, MobileNetV1 \cite{howard2017mobilenetsefficientconvolutionalneural} and MobileNetV2 \cite{sandler2019mobilenetv2invertedresidualslinear}, MobileNetV3 \cite{howard2019searchingmobilenetv3} incorporates numerous improvements. These include the addition of the Squeeze-and-Excitation (SE) module, the use of the Swish activation function, and optimizations in the network's tail structure. These enhancements allow MobileNetV3 to maintain its lightweight nature while also improving accuracy, making it an ideal fit for our research requirements.

\subsubsection{SE Module (Squeeze-and-Excitation)}
The SE module performs weighted operations on the feature maps output by each channel. It explicitly establishes interdependencies between feature channels and, through learning, computes the importance of each channel. Based on this importance, it assigns weights to the features of each channel, thereby highlighting important features and suppressing less important ones.

\paragraph{Squeeze Operation:}
The Squeeze operation aggregates global spatial information for each channel into a single value, effectively capturing the global context of the feature map. The mathematical expression is as follows:
\[
z_c = \frac{1}{H \times W} \sum_{i=1}^{H} \sum_{j=1}^{W} x_c(i,j),
\]
where $ H $ and $ W $ are the height and width of the feature map, respectively, and $ x_c(i,j) $ represents the value at position $(i,j)$ in channel $ c $.

\paragraph{Excitation Operation:}
After obtaining $ z $, we use a two-layer fully connected (FC) network to transform this information and learn the importance weights for each channel. The mathematical expression is as follows:
\[
s_c = \sigma(W_2 \cdot \delta(W_1 \cdot z)),
\]
where:
\begin{itemize}
    \item $ \delta $ represents the ReLU activation function,
    \item $ \sigma $ is typically the sigmoid function, which generates the final channel attention weights,
    \item $ W_1 $ and $ W_2 $ are the weight matrices of the two FC layers.
\end{itemize}

\paragraph{Final Output:}
The final output is obtained by multiplying the original feature map with the channel attention weights. The mathematical expression is:
\[
\tilde{x}_c = s_c \odot x_c,
\]
where:
\begin{itemize}
    \item $ \tilde{x}_c $: The weighted feature map for channel $ c $,
    \item $ \odot $: Element-wise multiplication.
\end{itemize}

This process ensures that important features are emphasized while less important ones are suppressed, improving the representational power of the model.

\subsubsection{Swish Activation Function}
The Swish activation function is an important innovation in MobileNetV3. It has proven to be more effective than the traditional ReLU activation function, especially in deeper neural networks. The smooth nature of the Swish function helps mitigate issues such as vanishing gradients and exploding gradients, thereby accelerating the training process and improving the convergence speed of the neural network. Particularly in deep neural networks with many layers, the Swish activation function facilitates better gradient propagation, preventing the vanishing gradient problem in earlier layers.

The mathematical expression for the Swish activation function is as follows:
\[
f(x) = x \cdot \sigma(x),
\]
where $ \sigma(x) $ is the standard Sigmoid function, defined as:
\[
\sigma(x) = \frac{1}{1 + e^{-x}}.
\]

This formulation combines the input $ x $ with its own sigmoid activation, allowing for a smooth, non-monotonic activation that improves performance in deep learning models.

\section{System Design and Implementation}

\subsection{Dataset Preparation}
We categorized the collected X-ray images related to tuberculosis into two classes: "Tuberculosis" and "Normal." To ensure that the model can generalize across different imaging conditions and patient populations, we employed a multi-level data augmentation strategy. First, all images were normalized by scaling their pixel values to the range $[0, 1]$, eliminating brightness variations caused by different acquisition devices. Next, we applied random cropping to resize the original images to a fixed size of $224 \times 224$ pixels, with reflection padding used at the boundaries to preserve important features. Additionally, horizontal flipping was randomly applied to the training samples to increase data diversity and reduce the risk of overfitting. These preprocessing steps not only enhanced the robustness of the model but also ensured consistency in the input data, laying a solid foundation for subsequent model training.

\subsection{Model Training and Optimization Strategy}
In the model training process, we selected the Softmax cross-entropy loss function as the optimization objective. Its mathematical expression is as follows:
\[
L = -\frac{1}{N} \sum_{i=1}^{N} \sum_{j=1}^{C} y_{ij} \log(p_{ij}),
\]
where $ N $ is the batch size, $ C $ is the number of classes, $ y_{ij} $ is the ground truth label, and $ p_{ij} $ is the predicted probability.

To accelerate model convergence and improve training stability, we adopted the Momentum Optimizer. Its mathematical expression is as follows:
\[
v_t = \beta v_{t-1} + \eta \nabla L(\theta),
\]
\[
\theta_t = \theta_{t-1} - v_t,
\]
where $ \beta $ is the momentum coefficient (set to 0.9), $ \eta $ is the learning rate (set to 0.001), and $ \nabla L(\theta) $ is the gradient of the loss function. During training, we employed a step-decay learning rate strategy, multiplying the learning rate by 0.9 every 10 epochs to ensure stable convergence to the optimal solution.

\subsection{Evaluation Metrics}
We evaluated the model using metrics such as accuracy, precision, recall, F1-score, and the area under the ROC curve (AUC). Visualization tools included confusion matrices, ROC curves, and heatmaps to comprehensively analyze model behavior. The mathematical expressions for these metrics are as follows:

\textbf{Accuracy:}
\[
{Accuracy} = \frac{\text{TP} + \text{TN}}{\text{TP} + \text{TN} + \text{FP} + \text{FN}},
\]

\textbf{Precision:}
\[
{Precision} = \frac{\text{TP}}{\text{TP} + \text{FP}},
\]

\textbf{Recall:}
\[
{Recall} = \frac{\text{TP}}{\text{TP} + \text{FN}},
\]

\textbf{F1-Score:}
\[
{F1-Score} = 2 \cdot \frac{\text{Precision} \cdot \text{Recall}}{\text{Precision} + \text{Recall}},
\]

\textbf{AUC:}
The AUC measures the ability of the model to distinguish between classes across all decision thresholds.

To visually demonstrate the diagnostic capability of the model, we used a confusion matrix to present the distribution of true positives (TP), false positives (FP), true negatives (TN), and false negatives (FN). Additionally, by plotting the ROC curve, we observed the trade-off between sensitivity and specificity at different decision thresholds. Heatmaps were used to visualize the metrics in the classification report, helping researchers quickly identify the strengths and weaknesses of the model. These diverse evaluation tools collectively formed a comprehensive performance analysis framework, providing reliable guidance for continuous model optimization.

\section{Results and Discussion}
\subsection{Performance Analysis}
\begin{figure}[h!]
    \centering
    \begin{minipage}{0.32\textwidth}
        \centering
        \includegraphics[width=\linewidth, height=4cm]{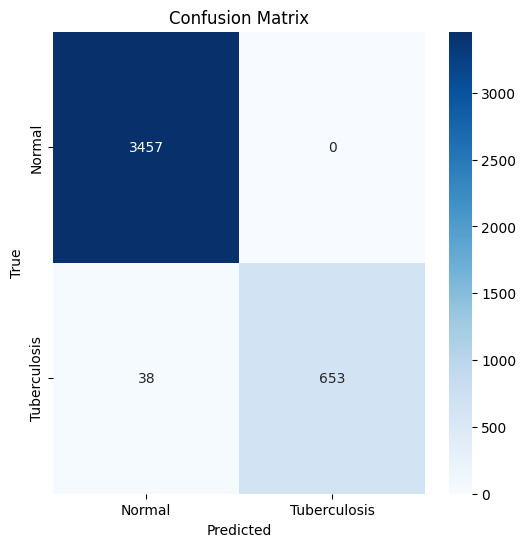} 
        \caption{Confusion Matrix (based on MindSpore)}
        \label{fig:confusion_matrix}
    \end{minipage}
    \hfill
    \begin{minipage}{0.32\textwidth}
        \centering
        \includegraphics[width=\linewidth, height=4cm]{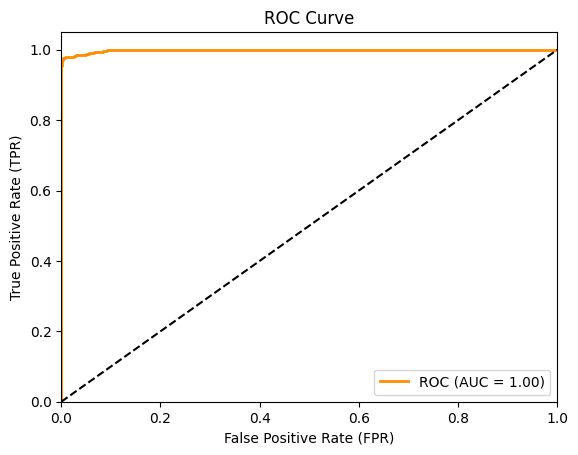} 
        \caption{ROC Curve (based on MindSpore)}
        \label{fig:roc_curve}
    \end{minipage}
    \hfill
    \begin{minipage}{0.32\textwidth}
        \centering
        \includegraphics[width=\linewidth, height=4cm]{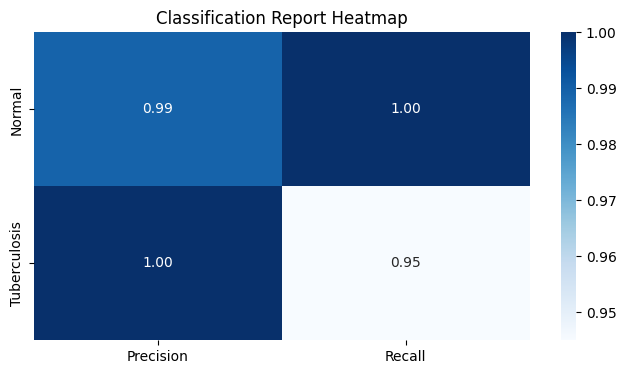} 
        \caption{Classification Report Heatmap (based on MindSpore)}
        \label{fig:heatmap}
    \end{minipage}
\end{figure}

After a rigorous training and optimization process, the developed tuberculosis diagnosis system demonstrated outstanding performance. In an independent test set containing 4,148 chest X-ray images, the model achieved an overall accuracy of 99.1\%, significantly surpassing traditional manual diagnostic methods. Through detailed analysis of the confusion matrix (Figure~\ref{fig:confusion_matrix}), we observed that the model performed exceptionally well in distinguishing between tuberculosis and normal samples: the true positive rate reached 94.5\%, the true negative rate was 100\%, while the false positive and false negative rates were 0\% and 5.5\%, respectively. This high level of classification accuracy is of great clinical significance as it directly impacts patients' subsequent treatment plans and health management.

The ROC curve (Figure~\ref{fig:roc_curve}) further confirmed the model's superior performance, with an area under the curve (AUC) of 0.99, close to the theoretical optimal value of 1.0. This metric indicates that, regardless of the decision threshold chosen, the model maintains an excellent balance between sensitivity and specificity. Particularly in critical regions, the model demonstrated strong discriminative ability, accurately distinguishing subtle pathological features. The classification report heatmap (Figure~\ref{fig:heatmap}) shows that the model maintained high consistency across all evaluation metrics: the precision for the "Normal" category was 99\%, with a recall of 100\%; for the "Tuberculosis" category, the precision was 100\%, and the recall was 95\%. These values fully demonstrate the system's stability and reliability.

\subsection{Edge Device Prediction Results}
\begin{figure}[h!]
    \centering
    \includegraphics[width=0.7\textwidth]{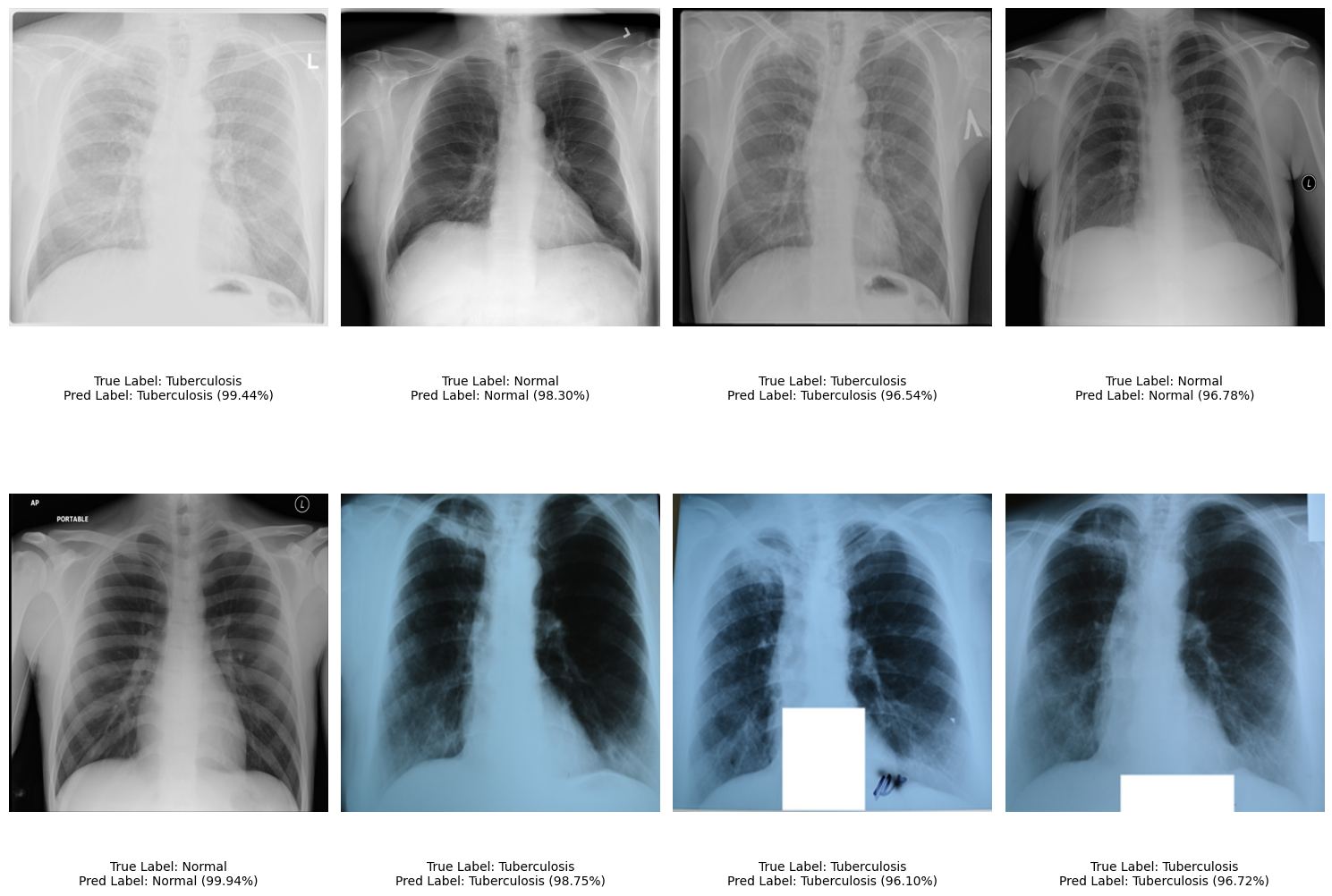}
    \caption{Prediction Results (based on MindSpore)}
    \label{fig:predictions}
\end{figure}

We conducted predictions on the 4,148 X-ray cases in the test set and randomly selected eight prediction results for display. As shown in Figure~\ref{fig:predictions}, the model's predictions perfectly matched the true labels without any misjudgments or missed diagnoses. This result fully validates the effectiveness and accuracy of the tuberculosis X-ray auxiliary diagnosis solution based on the Ascend 310 chip and MobileNetV3 model. In Figure~\ref{fig:predictions}, each image clearly displays the predicted class (tuberculosis or normal) along with the corresponding confidence score. High confidence scores indicate that the model has a strong grasp of its predictions, further enhancing our confidence in its performance. Particularly in areas with limited medical resources, this high-precision, low-cost auxiliary diagnosis solution will significantly improve early tuberculosis diagnosis rates and reduce the risks of misdiagnosis and missed diagnosis, thereby providing patients with more timely and effective treatment.

\subsection{Cost-Benefit Analysis and Practical Application Value}
From an economic perspective, the tuberculosis diagnosis system developed in this study demonstrates significant cost advantages. The entire system's hardware components include the Orange Pi AI Pro development board (integrated with the Ascend 310 chip) and related accessories, with a total cost controlled under \$150. Considering the procurement costs of traditional high-end medical imaging equipment, which often exceed hundreds of thousands of dollars, this price point makes the system highly suitable for large-scale deployment in resource-constrained environments. More importantly, the system's operating costs are also low: the typical power consumption of the Ascend 310 chip is only 8W, and the entire development board consumes just 20W. This means that even in areas with unstable power supply, the system can be powered by renewable energy sources such as solar panels.

From a practical standpoint, the system's compact design and low power consumption provide great convenience for its application in various medical scenarios. The Orange Pi AI Pro is highly portable and can be easily integrated into existing medical workflows. Its ability to support 16-channel full HD video decoding allows the system to process multiple patients' imaging data simultaneously, significantly improving diagnostic efficiency. Additionally, the system's response time is controlled at the millisecond level, enabling real-time diagnosis, which is particularly important in emergency scenarios requiring quick decisions.

In terms of social benefits, the widespread adoption of this system is expected to significantly improve tuberculosis prevention and control in areas with limited medical resources. By providing accurate, fast, and affordable diagnostic services, it not only reduces the workload of professional doctors but also helps grassroots medical institutions establish comprehensive screening mechanisms. Especially in remote rural areas, this edge computing solution can effectively address the shortage of professional talent, providing local residents with timely and reliable medical services. Moreover, the system's low-cost nature makes it easier to secure funding from governments and non-profit organizations, facilitating the global promotion of tuberculosis prevention and treatment efforts.

\section{Conclusion}
This study successfully developed and validated a low-cost, high-precision tuberculosis diagnosis system based on the MindSpore framework, Huawei Ascend 310 chip, and Orange Pi AI Pro device. By adopting the MobileNetV3-Large model, we achieved state-of-the-art diagnostic performance on low-cost, low-power hardware, with a test set accuracy of 99.1\% and an AUC value of 0.99, while keeping the total system cost under \$150. This achievement not only demonstrates the significant potential of edge computing technology in the field of medical diagnostics but also provides a practical solution to address diagnostic challenges in areas with limited medical resources.

However, the current system still has some limitations that need to be addressed in future research. First, in regions with poor medical infrastructure, the lack of X-ray equipment may limit the application scope of the system to areas where X-ray devices are available. Second, while the system performs exceptionally well in binary classification tasks, its ability to recognize other lung-related conditions, such as nodules, has not been fully validated. Finally, the system's user interface and interactive experience require further optimization to better adapt to the operational habits of grassroots medical personnel.

To address these challenges, future research directions include the following aspects: 
1. **Expand Dataset Scale and Diversity**: Collect more X-ray images of various lung lesions to improve the model's generalizability. 
2. **Explore Transfer Learning**: Leverage pre-trained models' knowledge transfer capabilities to accelerate model adaptation in new scenarios and enhance the accuracy of identifying rare cases. 
3. **Develop Multi-task Learning Frameworks**: Enable the system to simultaneously perform multiple related tasks, such as tuberculosis detection, lesion localization, size estimation, and severity assessment, providing more comprehensive clinical decision support. 
4. **Optimize User Interface**: Design more intuitive operational workflows and incorporate real-time feedback and interpretability features to enhance the system's usability and credibility.

We believe that with the implementation of these improvements, AI-assisted diagnostic systems based on edge computing will play an increasingly important role in global healthcare systems. Particularly in resource-constrained areas, this low-cost, high-precision solution has the potential to significantly improve early screening and diagnosis of tuberculosis, ultimately enhancing patients' health outcomes and quality of life.

\section*{Acknowledgments}
We would like to express our gratitude to the MindSpore Community for their invaluable support and resources throughout this research. Thanks for the support provided by MindSpore Community.

\bibliographystyle{unsrt}  


\end{document}